\documentclass[nohyper,12pt,a4paper]{JHEP3}
\usepackage{epsfig}
\usepackage{psfrag}
\usepackage{graphicx}
\usepackage{amsfonts,amsmath,amssymb}

\title{Black Hole Thermodynamics and Heavy Fermion Metals}

\author{E. J. Brynjolfsson,$^{1)}$ U. H. Danielsson,$^{2)}$, L. Thorlacius,$^{1),3)}$
and T. Zingg$^{1),3)}$\\
1) University of Iceland, Science Institute \\
Dunhaga 3, IS-107 Reykjavik, Iceland\\
\ \\
2) Institutionen f\"or fysik och astronomi\\
Uppsala Universitet, Box 803, SE-751 08 Uppsala, Sweden\\
\ \\
3) NORDITA, Roslagstullsbacken 23 \\
SE-106 91 Stockholm, Sweden \\ 
\ \\
E-mail: \email{erlingbr@hi.is}, \email{ulf.danielsson@physics.uu.se}, 
\email{lth@nordita.org}, \email{zingg@nordita.org}}

\abstract{Heavy fermion alloys at critical doping typically exhibit non-Fermi-liquid 
behavior at low temperatures, including a logarithmic or power law rise in the ratio 
of specific heat to temperature as the temperature is lowered. Anomalous specific 
heat of this type is also observed in a simple class of gravitational dual models 
that exhibit anisotropic scaling with dynamical critical exponent $z>1$.}


\preprint{NORDITA-2010-15, UUITP-09/10}

\begin{document}


\section{Introduction}

Heavy fermion systems exhibit very interesting thermodynamic properties 
at low temperature. The Sommerfeld ratio of the specific heat to temperature at
low $T$ is very large in these materials indicating an effective mass for the 
electrons near the Fermi surface that can be orders of magnitude larger than the
free electron mass, hence the term heavy fermions. In low-temperature experiments 
on certain heavy fermion alloys at critical doping the Sommerfeld ratio does not settle 
down to a constant, as it would for a conventional Fermi liquid, but continues to rise 
as the temperature is lowered \cite{Lohneysen:1994}. This behavior has also been
observed in systems at near-critical doping by tuning external parameters such as 
pressure or magnetic field \cite{Lohneysen:1994,Lohneysen:1996}. 
The non-Fermi-liquid behavior
of the specific heat is attributed to strongly correlated physics, governed by an 
underlying quantum critical point 
(see \cite{Stewart:1984zz,Stewart:2001zz,Lohneysen:2007zz} for reviews of 
heavy fermion systems) but a detailed understanding of quantum critical metals 
remains a key problem in theoretical physics \cite{Coleman:2005}. 

The last few years have seen increased interest in developing dual gravity 
models for strongly interacting systems in condensed matter physics 
(see \cite{Hartnoll:2009sz,Herzog:2009xv,Horowitz:2010gk,
McGreevy:2009xe,Sachdev:2010ch} for recent reviews). While the gravitational
approach has found a number of interesting applications it should be 
emphasized that its ultimate relevance to real condensed matter systems remains 
speculative. The original AdS/CFT conjecture \cite{Maldacena:1997re}, for which
there is by now strong evidence, relates maximally supersymmetric 3+1 dimensional 
Yang-Mills gauge theory with a large number of colors to supergravity on a ten
dimensional AdS$_5\times$S$_5$ spacetime. In condensed matter
physics one is instead interested in systems without supersymmetry where the
validity of a gravitational dual description is a more open question. Furthermore,
the large number of colors plays a key role in the supersymmetric gauge theory
context in justifying the use of large-scale classical 
geometry on the gravitational side of the duality and it is at present unclear what
the corresponding formal limit would be for the strongly coupled condensed 
matter systems that one wants to model. These are important issues to settle
but in the present paper we pursue a more modest goal of identifying yet 
another physical effect that can be phenomenologically modeled by a relatively 
simple dual gravitational system.

Gravity duals have one or more additional spatial dimensions compared to the
field theory in question.
Finite temperature effects are studied by having a black hole in the
higher-dimensional spacetime and when the black hole is electrically 
charged the Hawking temperature and the black hole charge provide
competing energy scales that can lead to interesting dynamics.
We will consider quantum critical points in three spatial dimensions that are 
characterized by a scaling law, 
\begin{equation}
t\rightarrow\lambda^{z}t,\qquad \vec{x}\rightarrow\lambda \vec{x}\,,
\label{lscaling}
\end{equation}
where the dynamical critical exponent $z$ is in general different from $1$. 
We are particularly interested in quantum phase transitions in heavy fermion 
alloys where the metal goes from being paramagnetic to being antiferromagnetic 
as the level of doping or some other external control parameter is varied 
\cite{Lohneysen:1994,Lohneysen:1996}. Theoretical work based on a lattice 
Kondo model suggests that the associated quantum critical point
exhibits anisotropic scaling with a non-universal dynamical critical exponent
that can be fitted by comparison to experimental data \cite{Si:2001}. 
In particular, it was found that $z\approx 2.7$ for the critically 
doped heavy fermion alloy CeCu$_{5.9}$Au$_{0.1}$, whose anomalous 
specific heat was reported in \cite{Lohneysen:1994}. 

In \cite{Kachru:2008yh} it was
conjectured that a strongly coupled system at a fixed point governed by
anisotropic scaling of this form could be modeled by a gravity theory in a
so called Lifshitz spacetime with the metric,\footnote{See \cite{Koroteev:2007yp} 
for early work on gravitational backgrounds with anisotropic scaling.}
\begin{equation}
ds^2=L^2\left(  -r^{2z}dt^2+r^2d\vec{x}^2+\frac{dr^2}{r^2}\right)  ,\label{lmetric}
\end{equation}
which is invariant under the transformation
\begin{equation}
t\rightarrow\lambda^{z}t,\quad 
\vec{x}\rightarrow\lambda \vec{x}, \quad
r\rightarrow\frac{r}{\lambda}\,.
\label{lifshitzscaling}
\end{equation}
Here $L$ is a characteristic length scale, which will determine the magnitude of 
the negative cosmological constant in the gravity model, and $(t,\vec{x},r)$ are 
dimensionless coordinates in the higher dimensional spacetime. 
The finite temperature physics of such a system is then encoded into the
geometry of a black hole, which is asymptotic to the Lifshitz spacetime (\ref{lmetric}).

In this paper we continue our study of asymptotically Lifshitz black holes
from \cite{Danielsson:2009gi,Brynjolfsson:2009ct} placing an emphasis
on their thermodynamic properties. We compute the specific heat of charged
asymptotically Lifshitz black holes as a function of temperature, keeping the
charge density in the dual field theory fixed. At high $T$ the thermodynamic 
behavior of the dual system is governed by the symmetries of the quantum 
critical point leading to a temperature dependence of the specific heat of
the form $C\sim T^{3/z}$. This is a direct consequence of scaling and we 
find the same result in the gravity dual using black hole thermodynamics.

At low $T$, we observe a crossover in the black hole thermodynamics indicating
non-trivial collective effects in the dual field theory. For $z=1$ our system reduces
to the thermodynamics of the well-known AdS-RN black branes. In this case,
the specific heat divided by $T$ goes to a constant value in the limit of low temperature, 
which is the same behavior as seen in a conventional Fermi liquid. This result was 
obtained previously in \cite{Rey:2008zz}, where it was interpreted as evidence for
the presence of a Fermi surface in the dual system. It was noted in \cite{Rey:2008zz}, 
however, from a holographic computation of transport properties that the 
low-energy effective theory near the Fermi surface cannot be that of weakly interacting 
fermions. Non-Fermi-liquid behavior has also been observed in spectral functions of 
probe fermions coupled to the $z=1$ system \cite{Liu:2009dm,Faulkner:2009wj}.

For a non-trivial dynamical critical exponent $z>1$ the black brane thermodynamics
calculation is more involved and we have to rely more heavily on numerical 
computations.  Interestingly, we find qualitatively different behavior in this case. 
The Sommerfeld ratio $C/T$ now continues to rise as we go to lower temperatures, 
which is indeed the behavior observed in critically doped heavy fermion systems. 
The ability of gravity models with Lifshitz scaling to capture this non-Fermi-liquid aspect 
of real quantum critical metals is interesting and is the main result of the present paper. 

\section{The holographic model}

The application that we have in mind involves a dual field theory 
in three spatial dimensions and accordingly the gravity dual is defined on a 4+1 
dimensional spacetime to accommodate an emergent extra dimension. Most of
our results carry over to other dimensions in a straightforward fashion. In particular,
similar conclusions can be drawn about planar systems with $d=2$.

We work with a $4+1$ dimensional version of the  holographic model of 
Kachru {\it et al.} \cite{Kachru:2008yh}, as formulated by Taylor \cite{Taylor:2008tg},
and with a Maxwell gauge field added, 
as in \cite{Brynjolfsson:2009ct}.
The bulk classical action for the gravitational sector consists of two parts,
\begin{equation}
S  = S_\textrm{Einstein-Maxwell} + S_\textrm{Lifshitz} \,.
\label{action}
\end{equation}
The first term is the usual action of $4+1$ dimensional 
Einstein-Maxwell gravity with a negative cosmological constant,
\begin{equation}
S_\textrm{Einstein-Maxwell} = \int\mathrm{d}^{5}x\sqrt{-g}\;\left(  
R-2\Lambda-\frac{1}{4}F_{\mu\nu}F^{\mu\nu}\right) .
\label{einsteinmaxwell}
\end{equation}
This is followed by a term involving a massive vector field,
\begin{equation}
S_\textrm{Lifshitz} = - \int\mathrm{d}^{5}x\sqrt{-g}\;
\left(\frac{1}{4} \mathcal{F}_{\mu\nu}\mathcal{F}^{\mu\nu} 
+\frac{c^2}{2}\mathcal{A}_\mu\mathcal{A}^\mu\right) ,
\label{lifshitzaction}
\end{equation}
whose sole purpose is to provide backgrounds with anisotropic scaling of the form 
(\ref{lifshitzscaling}). We refer to this auxiliary vector field, which only couples to gravity, 
as the Lifshitz vector field. 
The original gravity model of  \cite{Kachru:2008yh} was 3+1-dimensional and 
anisotropic scaling was obtained by including a pair of coupled two- and
three-form field strengths. Upon integrating out the three-form field strength, the remaining
two-form becomes a field strength of a massive vector and in this 
form the model is easily extended to general dimensions \cite{Taylor:2008tg}. 

The equations of motion obtained from the action (\ref{action}) consist of the Einstein 
equations, the Maxwell equations, and field equations for the Lifshitz vector,
\begin{eqnarray}
\label{einsteineqs}
G_{\mu\nu}+\Lambda g_{\mu\nu}&=&T_{\mu\nu}^\textrm{Maxwell}
+T_{\mu\nu}^\textrm{Lifshitz} \,, \\
\label{maxwelleqs}
\nabla_\nu F^{\nu\mu} &=&0 \,, \\
\label{lifshitzeqs}
\nabla_\nu \mathcal{F}^{\nu\mu} &=& c^2 \mathcal{A}^\mu \,, 
\end{eqnarray}
with the energy momentum tensors of the Maxwell and Lifshitz vector fields given by
\begin{eqnarray}
T_{\mu\nu}^\textrm{Maxwell}&=&
\frac{1}{2}(F_{\mu\lambda}F_{\nu}^{\>\lambda}
-\frac{1}{4}g_{\mu\nu}F_{\lambda\sigma}F^{\lambda\sigma}), \\
T_{\mu\nu}^\textrm{Lifshitz}&=&
\frac{1}{2}(\mathcal{F}_{\mu\lambda}\mathcal{F}_{\nu}^{\>\lambda}
-\frac{1}{4}g_{\mu\nu}\mathcal{F}_{\lambda\sigma}\mathcal{F}^{\lambda\sigma})
+\frac{c^2}{2}(\mathcal{A}_{\mu}\mathcal{A}_{\nu}
-\frac{1}{2}g_{\mu\nu}\mathcal{A}_{\lambda}\mathcal{A}^{\lambda}) .
\end{eqnarray}

The Lifshitz fixed point geometry (\ref{lmetric}) is a solution of the equations of
motion provided the characteristic length scale $L$ 
is related to the cosmological constant $\Lambda$ via
\begin{equation}
\Lambda=-\frac{z^{2}+2z+9}{2L^{2}}\,,
\label{cosmoconstant}
\end{equation}
the mass of the Lifshitz vector field is fine-tuned to $c=\sqrt{3z}/L$, and
the Lifshitz vector field has the background value
\begin{equation}
\mathcal{A}_t=\sqrt{\frac{2(z-1)}{z}}L\,r^z \,, \qquad 
\mathcal{A}_{x_i}=\mathcal{A}_r=0 \,.
\label{lbackground}
\end{equation}
The Maxwell field vanishes in the Lifshitz background, $A_\mu=0$.  

Although we do not do it here, it is straightforward to couple matter to this system.
Including a scalar field, for instance, leads to an instability at low $T$ giving rise
to holographic superconductors with Lifshitz scaling \cite{Brynjolfsson:2009ct}.
It is also interesting to include a coupling to a bulk Dirac spinor and obtain fermion 
spectral functions along the lines of \cite{Liu:2009dm,Faulkner:2009wj,Cubrovic:2009}.
We find that the spectral functions in the $z>1$ theory exhibit many of the same 
features as in the asymptotically AdS bulk spacetime considered in these earlier 
works, including peaks at certain values of momentum and frequency, but there 
are also important differences. In particular, the would be quasiparticle peaks are 
less sharply defined than their $z=1$ counterparts.\footnote{Work in progress.}

\section{Charged black branes }
In order to study finite temperature effects in the dual strongly coupled field theory we 
look for static black brane solutions of the equations of motion 
(\ref{einsteineqs}) - (\ref{lifshitzeqs}) which are asymptotic to the Lifshitz fixed point
solution (\ref{lmetric}). We consider a 4+1 dimensional metric of the form 
\begin{equation}
ds^2 =L^2 \left(-r^{2z}f(r)^2 dt^2+r^2d\vec{x}^2 +\frac{g(r)^2}{r^2}dr^2 \right) \,,
\label{metricansatz}
\end{equation}
for which the non-vanishing components of the vielbein can be taken to be
\begin{equation}
e_t^0=L\,r^z f(r)\,, \quad e_{x_1}^1 = \ldots =e_{x_3}^3 = L\,r\,, 
\quad e_r^{4}= L\,\frac{g(r)}{r} \,.
\label{vierbein}
\end{equation}
An asymptotically Lifshitz black brane with a non-degenerate horizon
has a simple zero of both $f(r)^2$ and $g(r)^{-2}$ at the horizon, which we take to be 
at $r=r_0$, and $f(r),\,g(r) \rightarrow 1$ as $r\rightarrow\infty$. 
It is straightforward to generalize this ansatz to include black holes with a spherical 
horizon or topological black holes with a hyperbolic horizon but it is the flat horizon 
case (\ref{metricansatz}) that is of direct interest for the gravitational dual description 
of a strongly coupled $3+1$ dimensional field theoretic system. 

In a static electrically charged black brane background the Maxwell gauge field and the 
Lifshitz vector can be chosen to be of the form
\begin{equation}
A_M= \{\alpha(r),0,\ldots,0\} \,; \qquad 
\mathcal{A}_M = \sqrt{\frac{2(z-1)}{z}} \{a(r),0,\ldots,0\} \,,
\end{equation}
where the corresponding coordinate frame components are given by  
$A_\mu=e_\mu^M A_M$ and 
$\mathcal{A}_\mu=e_\mu^M \mathcal{A}_M$. 

\subsection{Field equations for static configurations}

The Maxwell equations and the equations
of motion for the Lifshitz vector can be written in first-order form,
\begin{eqnarray}
\frac{r}{f}\frac{d}{dr}\left(f\, \alpha\right)&=& -z\,\alpha+g\,\beta  , \label{maxwell1} \\
r\frac{d}{dr}\left(r^3 \beta\right)&=& 0, \label{maxwell2} \\
\frac{r}{f}\frac{d}{dr}\left(f\, a\right)&=& -z\, a+z\, g\, b , \label{lifshitz1} \\
r\frac{d}{dr}\left(r^3 b\right)&=& 3 r^3 g\, a  \label{lifshitz2} ,
\end{eqnarray}
where $\beta$ and $b$ are defined via (\ref{maxwell1}) and (\ref{lifshitz1}) respectively.
The Maxwell equation  (\ref{maxwell2}) trivially integrates to
\begin{equation}
\beta(r) = \frac{\rho}{r^3}\,,
\end{equation}
where the integration constant $\rho$ is proportional to the world-volume electric charge 
density of the black brane, which in turn is identified with the charge density in the dual 
field theory by the standard AdS/CFT dictionary \cite{Gubser:1998bc,Witten:1998qj}, 
as explained in detail in \cite{Hartnoll:2007ai}.

The Einstein equations reduce to the following pair of first-order differential equations
\begin{eqnarray}
\frac{r}{g}\frac{dg}{dr}&=& 
\frac{g^2}{3}\left[\frac{(z-1)}{2}\left(3a^2+z b^2\right)-\frac{(z^2+2z+9)}{2}
+\frac{\rho^2}{4r^{6}} \right]+2\,, 
\label{einsteintt} \\
\frac{r}{f}\frac{df}{dr}&=& 
\frac{g^2}{3}\left[\frac{(z-1)}{2}\left(3a^2-z b^2\right)+\frac{(z^2+2z+9)}{2}
-\frac{\rho^2}{4r^{6}} \right]-z-1.
\label{einsteinrr} 
\end{eqnarray}
The Lifshitz fixed point solution is given by $\alpha=\rho=0$ and $f=g=a=b=1$. 

To bring out the underlying scale invariance of the model we rewrite
the field equations using $u=\log(r/r_0)$ instead of $r$ and introduce
the scale invariant ratio $\tilde{\rho}\equiv \rho/r_0^d$. The remaining 
Maxwell equation (\ref{maxwell1}) and the equations of motion of the Lifshitz field
(\ref{lifshitz1}) and (\ref{lifshitz2}) reduce to
\begin{eqnarray}
\dot{\alpha}+\frac{\dot{f}}{f} \alpha&=& -z \alpha+\tilde{\rho}e^{-3u}g\,, 
\label{maxwell1u} \\
\dot{a}+\frac{\dot{f}}{f}a&=& -z a+z g b \,, 
\label{lifshitz1u} \\
\dot{b}&=& -3 b+3 g a\,, 
\label{lifshitz2u} 
\end{eqnarray}
where $\dot{\ }\equiv\frac{d}{du}$. The Einstein equations 
(\ref{einsteintt}) and (\ref{einsteinrr}) can similarly be 
rewritten
\begin{eqnarray}
\frac{\dot{g}}{g}&=&  \frac{g^2}{6} \left[(z-1)\left(3a^2+zb^2\right)
+\frac{\tilde{\rho}^2}{2}e^{-6u} -(z^2+2z+9)
\right]+2, 
\label{einstein2} \\
\frac{\dot{f}}{f}&=& \frac{g^2}{6} \left[(z-1)\left(3a^2-zb^2\right)
-\frac{\tilde{\rho}^2}{2}e^{-6u} +(z^2+2z+9)
\right]-z-1. 
\label{einstein1} 
\end{eqnarray}
Using these variables, the field equations are manifestly invariant under the 
rescaling (\ref{lifshitzscaling}). For given $z\geq 1$, there is a one parameter
family of black brane solutions, labelled by $\tilde{\rho}$. A neutral black brane has
$\tilde{\rho}=0$ while the extremal limit is given by
$\vert \tilde{\rho}\vert\rightarrow \sqrt{2(z^2+2z+9)}$.

\subsection{Exact solutions}

When $z=1$ the terms involving $a$ and $b$ on the right hand 
side of the Einstein equations drop out. The equations then reduce to those of 
Einstein-Maxwell theory with a negative cosmological constant and one finds 
the AdS-RN solution describing charged black branes in $4+1$ dimensional 
asymptotically AdS spacetime,
\begin{eqnarray}
f^2=\frac{1}{g^2}&=&\left(1-e^{-2u}\right)
\left(1+e^{-2u}-\frac{\tilde{\rho}^2}{12}e^{-4u}\right),\\
A_t&=&\frac{L r_0\tilde{\rho}}{2}\,\left(1-e^{-2u}\right).
\label{adsrn}
\end{eqnarray} 
At $z>1$ there is a corresponding family of charged asymptotically Lifshitz black 
branes. In general these solutions can only be obtained numerically, but it turns out 
that in the special case $z=6$ an isolated exact solution can be found 
\cite{Brynjolfsson:2009ct,Pang:2009pd},
\begin{equation}
b=1\,, \quad f^2=\frac{1}{g^2}=a^2=1-e^{-6u}, \quad 
A_t=\pm\sqrt{2}L r_0^{6}\,(e^{3u}-1).
\label{z6exact}
\end{equation}
We will primarily rely on numerical solutions in 
the following but the AdS-RN solution and the exact $z=6$ solution 
are useful for checking the numerics.

\subsection{Numerical solutions}

The family of black brane solutions at generic $z> 1$ can be mapped out using 
numerical techniques similar to those of \cite{Danielsson:2009gi}. 
The numerical integration is started near the 
black hole, with suitable boundary conditions for a regular non-degenerate 
horizon at $u=0$, and then proceeds outwards towards the asymptotic region. 
First of all, there should be a simple zero of $f^2$ and $g^{-2}$ at the horizon.
Furthermore, the product $f \alpha$ must have a simple zero at the horizon in 
order for the gauge connection to be regular there. Similarly, the combination $g a$ 
on the right hand side of (\ref{lifshitz2u}) should be regular at the horizon and it then
follows from (\ref{lifshitz1u}) that $b$ takes a finite value there. Putting all of this 
together we find that the near-horizon behavior can be parametrized as follows
\begin{eqnarray}
f(u)&=&\sqrt{u}(f_0+f_1 u+f_2 u^2+\ldots), \label{fexpansion} \\
g(u)&=&\frac{1}{\sqrt{u}}(g_0+g_1 u+g_2 u^2+\ldots), \label{gexpansion} \\
\alpha(u)&=&\sqrt{u}(\alpha_0+\alpha_1 u+\alpha_2 u^2+\ldots), 
\label{alphaexpansion} \\
a(u)&=&\sqrt{u}(a_0+a_1 u+a_2 u^2+\ldots), \label{aexpansion} \\
b(u)&=&b_0+b_1 u+b_2 u^2+\ldots . \label{bexpansion} 
\end{eqnarray}
Inserting these near-horizon expansions into the equations of motion we obtain
a two-parameter family of initial value data for any given $z$, parametrized for instance 
by $\tilde{\rho}$ and $b_0$. It turns out, however, that these two parameters cannot be 
varied independently but are restricted by the condition that the metric and Lifshitz vector 
approach the Lifshitz fixed point solution (\ref{lmetric}) and 
(\ref{lbackground}) sufficiently rapidly as $u\rightarrow \infty$ 
\cite{Danielsson:2009gi,Bertoldi:2009vn,Ross:2009ar}. 
This means that for a given value of $z$ the parameter $\tilde{\rho}$ uniquely determines 
a charged black brane solution, up to overall scale.

\subsection{Asymptotic behavior at large $u$}
\label{largeuasymptotics}

We now consider the asymptotic behavior of our fields as $u\rightarrow\infty$. The 
first observation is that the system of equations (\ref{maxwell1u}) - (\ref{einstein1}) only 
contains the combination $\frac{\dot{f}}{f}$ and therefore a uniform rescaling of $f$ is a 
symmetry of the equations. The symmetry is fixed, however, in an asymptotically Lifshitz 
solution for which $f\rightarrow 1$ as $u\rightarrow\infty$.\footnote{When solving the 
equations numerically, it is convenient to initially put $f_0=1$ in the near-horizon 
expansion (\ref{fexpansion}) when setting up the numerical integration and then 
rescale $f$ at the end of the day so that $f\rightarrow 1$ as $u\rightarrow\infty$.}

We next observe that (\ref{einstein1}) can be used to eliminate $f$ from 
(\ref{lifshitz1u}) -- (\ref{einstein2}) leaving a closed system of 
first-order equations for $a$, $b$, and $g$. The large $u$ behavior can be obtained by
linearizing around the Lifshitz fixed point $a=b=g=1$. The corresponding problem
for $d=2$ was discussed in \cite{Danielsson:2009gi,Bertoldi:2009vn} and we omit the 
details here. 
At large $u$, solutions of the full non-linear system 
approach a linear combination of the eigenmodes of the linearized system
plus a universal mode coming from source terms involving $\tilde{\rho}^2$. 
The leading large $u$ behavior of the eigenmodes is $O(e^{\lambda_i u})$ with 
the eigenvalues $\lambda_i\in\{-3-z,\frac{1}{2}\left(-3-z\pm\sqrt{9z^2-26z+33}\right)\}$ 
while the universal mode falls off as $O(e^{-6 u})$ independent of $z$.

\FIGURE{\epsfig{file=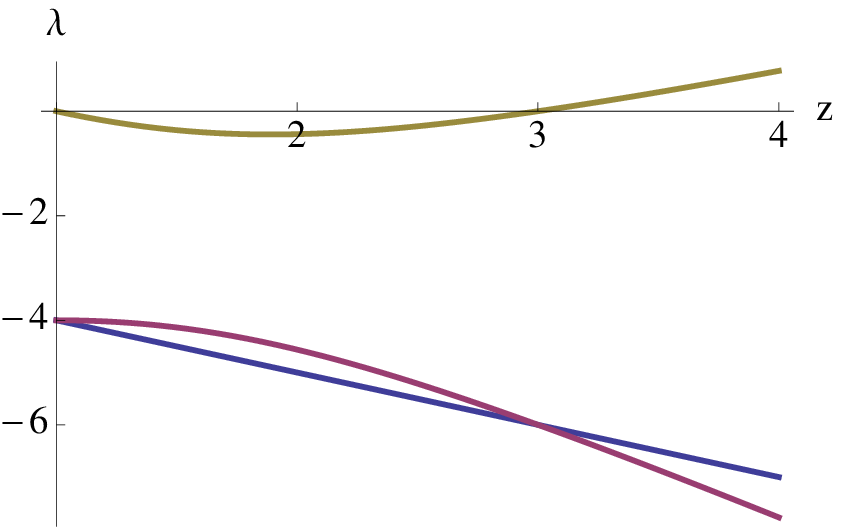,width=6cm} 
\caption{Eigenvalues of the
linearized problem for $\{g,b,a\}$ as a function of $z$ for $d=3$.} \label{eigenv}}

Figure~\ref{eigenv} shows the three eigenvalues as a function of $z$. 
The eigenmode that belongs to the largest eigenvalue
is problematic \cite{Danielsson:2009gi,Bertoldi:2009vn,Ross:2009ar}. 
For $z\geq 3$ it is non-negative and a solution that contains it fails to be 
asymptotically Lifshitz. For $1<z<3$ it is a negative mode but the falloff at
large $u$ is too slow to give finite energy. This mode is eliminated in our
black hole solutions by fine-tuning the value of $b$ at the black hole
horizon. For $z<3$ the universal $e^{-6u}$ mode is sub-leading compared to 
the two remaining eigenmodes of the linearized system but at $z>3$ it 
dominates the asymptotic behavior. In the dividing $z=3$ case the linearized 
system is degenerate and the asymptotic behavior includes powers of
$u$ on top of the $e^{-6u}$ falloff. 

Finally, we turn to equation (\ref{maxwell1u}) for the Maxwell field. 
In the Lifshitz background with $f=g=1$ it has the solution
\begin{equation}
\alpha = \left\{ 
\begin{array}{lcc}
\tilde{\mu}\,e^{-zu}+\frac{\tilde{\rho}}{(z-3)}e^{-3u} 
& \quad\textrm{if}\quad & z\neq 3 \,, \\
\tilde{\mu}\,e^{-3u}+\tilde{\rho}\,u\,e^{-3u} 
& \quad\textrm{if}\quad & z= 3 \,.
\end{array}
\right.
\label{vpotential}
\end{equation}
The non-vanishing component of the vector potential in a coordinate frame is then
\begin{equation}
A_t = \left\{ 
\begin{array}{lcc}
\mu+\frac{L\rho}{(z-3)} r^{z-3}
& \quad\textrm{if}\quad & z\neq 3 \,, \\
\mu+L\rho\log{(\frac{r}{r_0})}
& \quad\textrm{if}\quad & z= 3 \,,
\end{array}
\right.
\label{Apotential}
\end{equation}
with $\mu=L\tilde{\mu}\,r_0^z$. 
Once again the qualitative behavior of solutions changes at $z=3$. For low
$z$ values, in the range $1<z<3$, the term involving the chemical potential 
$\mu$ dominates compared to the term involving the charge density
$\rho$ at large $r$, while at $z\geq 3$ it is the charge density term that
is leading \cite{Hartnoll:2009ns}. 

In general, we are not working with Lifshitz background itself but with
solutions of the full non-linear system of equations that are only asymptotically
Lifshitz. It turns out, however, the leading large $u$ behavior of $\alpha$ 
carries over from the Lifshitz background to more general case as 
long as the value of $z$ isn't too high.\footnote{At $z\geq 9$ we expect non-linear 
effects to give rise to additional terms in (\ref{vpotential}) with a falloff in
between that of the charge density and chemical potential terms. We will
not keep track of such terms here since all systems of physical interest that
we are aware of have $z< 9$.} 

\section{Black brane thermodynamics}
\label{branethermo}

So far, we have set up the holographic model and considered charged black
brane geometries that are conjectured to provide a dual description of a 
strongly coupled 3+1 dimensional system near a quantum critical point with 
dynamical critical exponent $z\geq 1$. These black branes turn out to have
interesting thermodynamics. At high temperature the thermodynamic behavior
is governed by the underlying Lifshitz symmetry, but collective effects come 
into play as the temperature is 
lowered and modify the thermodynamics.
To see this, we use a combination of analytic and numerical arguments to 
compute the specific heat of the system as a function of temperature. Our 
starting point is the following expression for the Hawking temperature of
a black brane,\footnote{From here on we fix the characteristic length
scale as $L=1$. Explicit factors of $L$ can be reintroduced into the formulas 
by dimensional analysis.}
\begin{equation}
T_H=\frac{r_0^z}{4\pi}\,\frac{f_0}{g_0}\,,
\label{hawkingtemp}
\end{equation}
which is obtained in the standard way by requiring the Euclidean metric
of the black brane to be smooth at the horizon at $r=r_0$. The coefficients
$f_0$ and $g_0$ are taken from the near-horizon expansions 
(\ref{fexpansion}) and (\ref{gexpansion}) in the scale invariant variable $u$
and only depend on the value of $\tilde{\rho}$. The temperature can therefore
be expressed as follows,
\begin{equation}
T_H=\frac{r_0^z}{4\pi}\,F_z(\tilde{\rho})\,,
\label{hawkingtemp2}
\end{equation}
with $F_z(\tilde{\rho})\equiv \frac{f_0}{g_0}$ for different values of $z$ 
shown in Figure~\ref{Ffigure}. 

\FIGURE{\epsfig{file=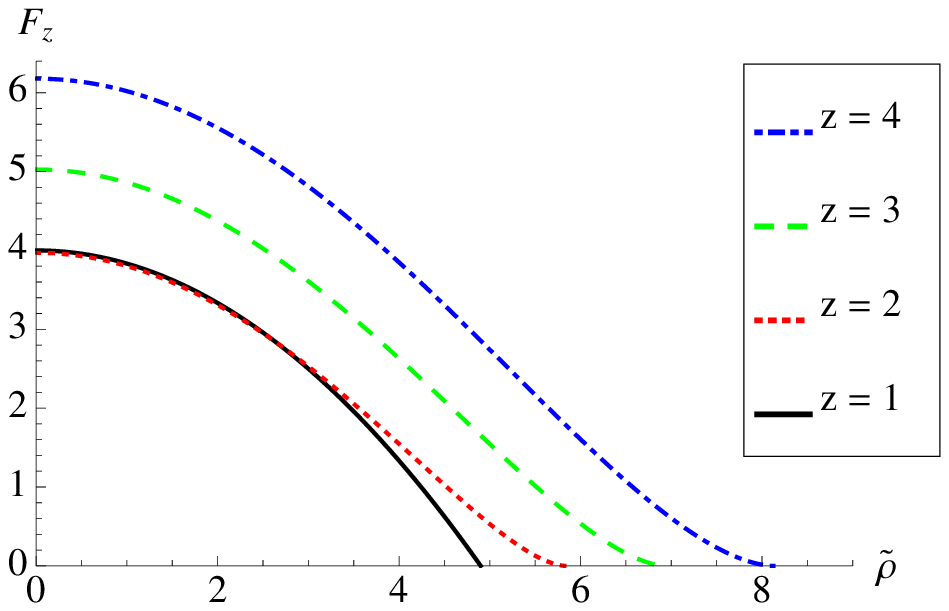,width=7.2cm} \caption{Temperature
function $F_z(\tilde{\rho})$ for several $z$ values. The $z=1$ curve plots
the exact result (\ref{f1func}) while the $z>1$ curves are obtained from 
numerical black hole solutions.}\label{Ffigure}}

The specific heat at fixed volume in the boundary theory is then 
\begin{equation}
C=T\,   \frac{dS}{dT}=T \frac{(dS/dr_0)}{(dT/dr_0)}\,,
\label{heatcapacity}
\end{equation}
where $S=\frac{1}{4} r_0^3$ is the Bekenstein-Hawking entropy density 
of the black brane and $T$ in the boundary system is identified with 
the Hawking temperature (\ref{hawkingtemp2}). 
We have to make a choice whether to work at fixed charge density $\rho$ or fixed
chemical potential $\mu$ in the boundary theory when calculating thermodynamic
quantities. For the field theory problem that we wish to model it is natural to keep the 
charge density fixed since the net density of charge carriers is given by the density of 
dopants, which is fixed in a given sample \cite{Hartnoll:2009ns}. In this case
\begin{equation}
 \frac{d}{dr_0}=\frac{\partial}{\partial r_0}-\frac{3\tilde\rho}{r_0}
\frac{\partial}{\partial \tilde\rho} \,,
\end{equation}
and we find
\begin{equation}
\frac{C}{T}=\frac{3\pi r_0^{3-z}}{zF_z(\tilde\rho)-3\tilde\rho F'_z(\tilde\rho)}.
\label{cTequation}
\end{equation}
If we instead were to keep the chemical potential $\mu$ fixed then 
\begin{equation}
 \frac{d}{dr_0}=\frac{\partial}{\partial r_0}-\frac{z\tilde\mu}{r_0}
\left(\frac{d \tilde\mu}{d\tilde\rho}\right)^{-1} \frac{\partial}{\partial \tilde\rho} \,,
\end{equation}
with $d\tilde\mu/d\tilde\rho$ obtained from the asymptotic behavior of the
Maxwell field (\ref{vpotential}) in the numerical black brane solution, and then
the remaining steps in the calculation of the specific heat are parallel to those
at fixed charge density. 

The variables $r_0$ and $\tilde{\rho}$ on the right hand side of equation 
(\ref{cTequation}) refer to the 4+1 dimensional black brane geometry while it is $T$ 
and $\rho$ that have direct interpretation in the dual field theory. The latter variables 
are expressed in terms of the former via the definition $\tilde{\rho}=\rho/r_0^3$ and 
equation (\ref{hawkingtemp2}) for the Hawking temperature. We can numerically invert 
these relations in order to express the specific heat (\ref{heatcapacity}) in terms $T$ 
and $\rho$. Figure~\ref{fig:variables} shows the map between brane variables and
physical variables for $z=2$ and similar maps are obtained for other $z$ values.

\FIGURE[ht]{\epsfig{file=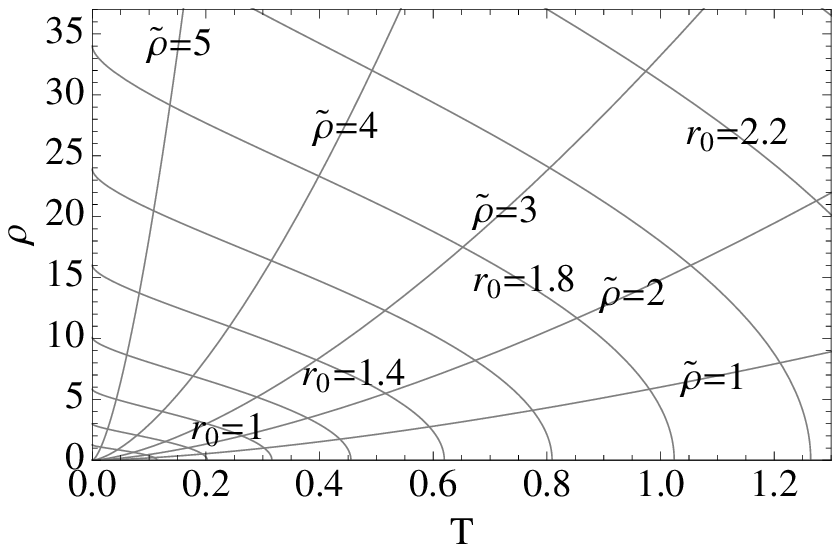,width=7.0cm} 
\caption{Contour map expressing $r_0$ and $\tilde{\rho}$ in terms of
$T$ and $\rho$ at $z=2$.}\label{fig:variables}}

\subsection{Scaling at high temperature}

Below, we present numerical results for $C/T$ obtained via the above procedure for
several values of $z$ but some information can be obtained by analytic arguments.
In particular, we can extract a scaling relation for the specific heat at high temperature. 
The argument is the same for any number of spatial 
dimensions $d$. Writing $\tilde{\rho}=\rho/r_0^d$, the expression
(\ref{hawkingtemp}) for the temperature remains the same except that the 
detailed shape of the function $F_z(\tilde{\rho})$ depends on $d$. 
The limit of high temperature at fixed $\rho$ corresponds to large $r_0$ and 
$\tilde{\rho}\rightarrow 0$. The field equations depend on $\tilde{\rho}^2$ in a 
smooth way so we expect $F_z(\tilde{\rho})$ to be a smooth even function of 
$\tilde{\rho}$ (this is also evident from the graphs in Figure~\ref{Ffigure}) and the 
denominator in the $d$ dimensional version of equation (\ref{cTequation}) 
reduces to $zF_z(0)$. 
As a result the temperature and the entropy density, and their first derivatives,
depend only on the overall 
scale $r_0$ in the high-temperature limit and the electric charge carried by the black
brane does not affect the dynamics to leading order,
\begin{equation}
T \approx \frac{F_z(0)}{4\pi}\,r_0^z\,,\qquad S=\frac{1}{4}r_0^d\, .
\label{hawkingtemp3}
\end{equation}  
The high-temperature behavior of the specific heat then immediately follows from 
the general expression (\ref{heatcapacity}),
\begin{equation}
C \sim r_0^d \sim T^{d/z},
\label{scalinglaw}
\end{equation}  
which reduces, in particular, to $C\sim T^{3/z}$ when $d=3$. It is straightforward 
to go a step further and integrate both sides of (\ref{heatcapacity}) with respect to 
$T$ to obtain the following relation between energy and entropy of the system 
in the high-temperature limit,
\begin{equation}
E = \frac{d}{d+z} T\,S \,,
\label{energy}
\end{equation}  
recovering the result found previously for electrically neutral black branes 
in \cite{Bertoldi:2009dt}. The scaling behavior can be traced to the underlying 
Lifshitz symmetry of the quantum critical theory.  It is easy 
to see that a generic statistical mechanical 
system in three spatial dimensions with a dispersion relation of the form 
$\omega \sim k^z$ exhibits the same scaling behavior \cite{Bertoldi:2009dt}.

\subsection{Low-temperature behavior}

At low temperature, the black brane thermodynamics exhibits 
interesting behavior due to collective effects in the dual field theory and the simple
scaling that is seen at high temperature no longer applies. The behavior is 
qualitatively different in conformal systems with $z=1$ as compared to Lifshitz 
systems with $z>1$ and we consider these cases in turn. 

At $z=1$ we have the exact AdS-RN black brane solution (\ref{adsrn}) 
for which the Hawking temperature is easily determined. One finds 
\begin{equation}
F_1(\tilde{\rho}) = 4-\frac{\tilde{\rho}^2}{6}\,,
\label{f1func}
\end{equation}
and the specific heat at fixed $\rho$ is given by
\begin{equation}
\frac{C}{T}=\frac{18\pi r_0^2}{24+5\tilde{\rho}^2} . 
\label{exactc}
\end{equation}
At low temperature the charge on the black brane approaches the extremal limit,
$\tilde{\rho}\rightarrow\sqrt{24}$ and the 
specific heat depends linearly on temperature,
\begin{equation}
\frac{C}{T}\rightarrow \frac{\pi\rho^{2/3}}{16\cdot 3^{1/3}} 
\quad \textrm{as}\quad T\rightarrow 0.
\end{equation}
A weakly-coupled Fermi liquid has a specific heat that is linear in $T$ at low 
temperatures and it is interesting to see the same behavior emerge in the low-temperature
limit of a black brane in Einstein-Maxwell theory without having made any reference 
to specific matter fields in the calculation. This result was obtained previously 
in \cite{Rey:2008zz} where it was taken as evidence for the existence of a Fermi surface 
in the dual field theory. Spectral functions of probe fermions coupled to the $z=1$
system computed in \cite{Liu:2009dm,Faulkner:2009wj,Cubrovic:2009} also strongly
suggest the presence of a Fermi surface. 
Having a specific heat that is linear in $T$ and a Fermi surface does not imply that the 
system consists of weakly coupled fermions. Indeed, the scaling behavior of excitations
near the Fermi surface differs from that of a Landau Fermi liquid in the probe fermion
analysis of \cite{Liu:2009dm,Faulkner:2009wj} and the zero-temperature conductivity
computed in \cite{Rey:2008zz} was found to be diffusive rather than ballistic, suggesting 
that disorder plays a role.

\FIGURE[ht]{\epsfig{file=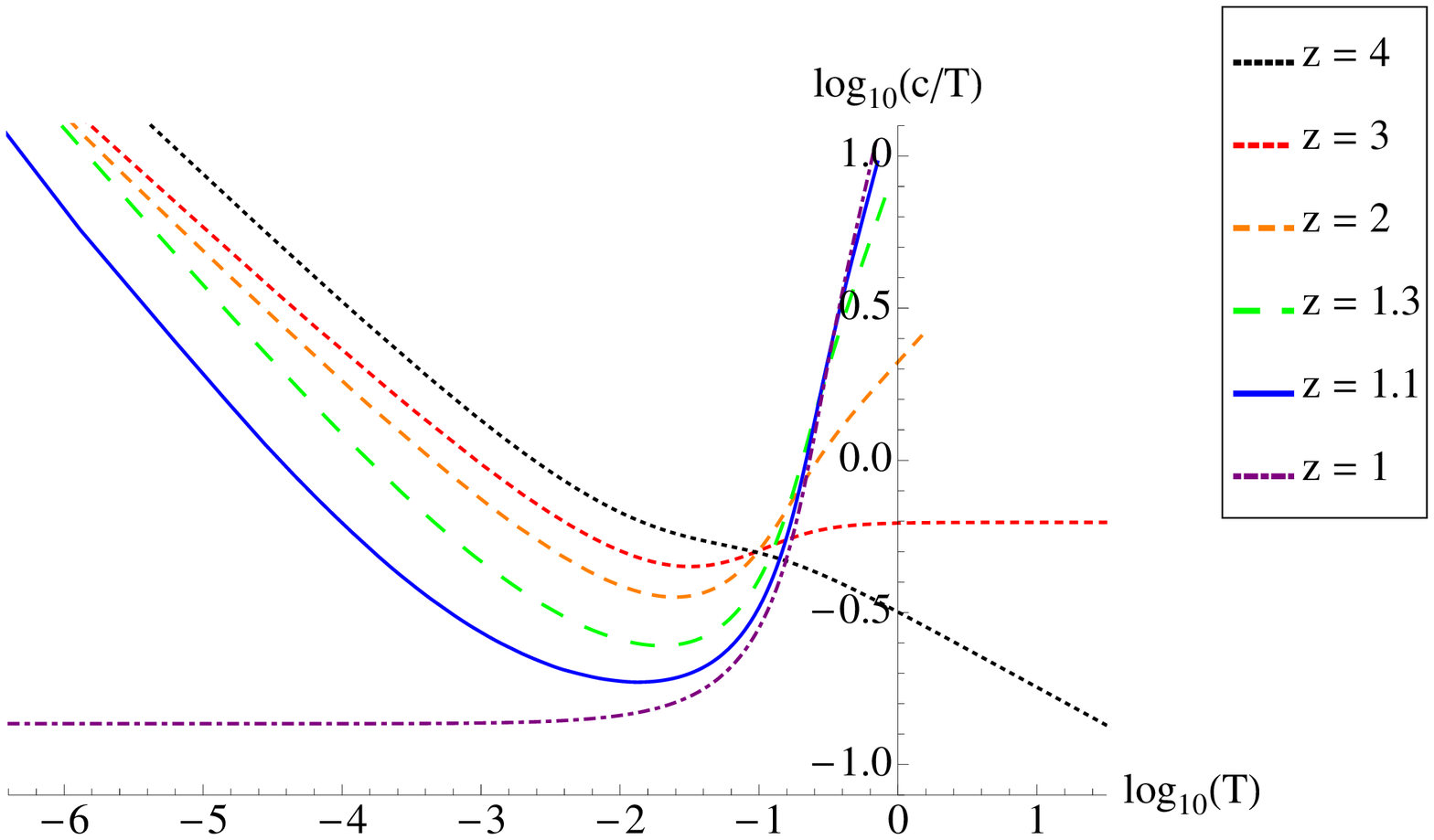,width=12.5cm} \caption{Black brane 
specific heat divided by temperature for several $z$ values calculated at fixed charged 
density $\rho=1$. The $z=1$ curve is obtained from the exact result (\ref{exactc}).}
\label{fig:coverT}}

At $z>1$ we do not have explicit analytic black brane solutions except the
isolated $z=6$ solution (\ref{z6exact}) and our results 
for the specific heat in $z>1$ systems are therefore based on numerical
black brane solutions with non-vanishing Lifshitz vector field.
The key qualitative difference compared to the $z=1$ case can be seen in 
Figure~\ref{Ffigure}. The function $F_1(\tilde{\rho})$ goes to zero in a linear
fashion in the extremal limit $\tilde{\rho}\rightarrow \sqrt{24}$ but for $z>1$ both 
$F_z(\tilde{\rho})$ and its first derivative go to zero in the
extremal limit $\tilde{\rho}\rightarrow \sqrt{2(z^2+2z+9)}$. It then follows from
equation (\ref{cTequation}) that the ratio $C/T$ diverges in the $T\rightarrow 0$ 
limit for $z>1$. The growth in $C/T$ towards lower temperatures is evident in 
the $z>1$ curves in Figure~\ref{fig:coverT}.\footnote{Finite numerical precision limits
how far the curves for $z>1$ in Figure~\ref{fig:coverT} can be extended towards 
low $T$.}
This trend is in qualitative 
agreement with the measured specific heat of certain heavy fermion 
alloys \cite{Stewart:2001zz}. The high $T$ behavior of the specific heat
curves in the figure, on the other hand, matches the scaling law 
(\ref{scalinglaw}).

\section{Discussion}
In this paper we have employed a relatively simple holographic
description of a strongly coupled 3+1 dimensional quantum critical point
with asymmetric scaling to model the anomalous specific heat found at
low temperature in many heavy fermion compounds. The calculation 
is performed entirely within the context of black hole thermodynamics and
the nature of the low-lying spectrum of excitations is thus investigated 
without introducing any specific matter probes. It would be very interesting
to complement the results obtained here by a study of spectral functions 
for probe fermions, generalizing the $z=1$ work 
of \cite{Liu:2009dm,Faulkner:2009wj,Cubrovic:2009}, and a calculation 
of conductivities. This requires us to extend existing gravitational techniques 
for obtaining spectral functions and transport coefficients to models exhibiting 
Lifshitz scaling with non-trivial dynamical critical exponent $z>1$, and our 
work in this direction is in progress. 

One limitation of the present work concerns the non-vanishing black hole
area in the extremal limit, which corresponds to having non-vanishing
entropy at zero temperature in the dual system. This can for instance be 
remedied by coupling the system to a scalar field and include the back-reaction 
to the scalar hair carried by the black hole at very low temperatures. 
In this case the area of the black hole horizon shrinks as the temperature is
lowered and the system no longer has finite zero-temperature entropy.
This was shown in \cite{Horowitz:2009ij} for asymptotically AdS black 
holes with hair and we have obtained analogous results in our $z>1$ 
model with scalar matter included \cite{Brynjolfsson:2010mk}. 
The scalar hair corresponds to having a charged superfluid condensate in the 
dual theory and the instability to developing superfluidity bears out the 
general expectation that scale invariant quantum critical matter is never 
found as the true ground state of a system. A further study of the interplay 
between the heavy fermion physics and superfluidity in these holographic 
models is an interesting avenue for further work.

We find it intriguing that experimentally measured deviations from Fermi liquid 
behavior in critically doped heavy fermion alloys can be qualitatively 
reproduced by holographic duals with Lifshitz symmetry. The curves for $z>1$ 
in Figure~\ref{fig:coverT} suggest a power law for $C/T$ at low temperature 
rather than a logarithm. The experimental data vary from one material to 
another and both logarithmic and power law fits are used \cite{Stewart:2001zz}. 
The fact that our curves for $z>1$ approach straight lines at low $T$ 
should not be over interpreted but it is encouraging to see the right
trend coming from simple gravitational models with Lifshitz scaling.

\acknowledgments

This work was supported in part by the G\"{o}ran Gustafsson foundation, the 
Swedish Research Council (VR), the Icelandic Research Fund, the University of 
Iceland Research Fund, and the Eimskip Research Fund at the University of Iceland.
We thank E. Ardonne, H. Hansson, and P. Kakashili for useful discussions.

\end{document}